# Interfacial viscosity-induced suppression of lateral migration of a surfactant-laden droplet in a non-isothermal Poiseuille flow

Devi Prasad Panigrahi[a], Somnath Santra[b], Theneyur Narayanaswamy Banuprasad[c], Sayan Das[b] and Suman Chakraborty[b, c]†

[a]*Department of Mechanical Engineering, Indian Institute of Technology Kanpur, Uttar Pradesh 208016, India*

[b]*Department of Mechanical Engineering, Indian Institute of Technology Kharagpur, West Bengal - 721302, India*

[c]*Advanced Technology Development Centre, Indian Institute of Technology Kharagpur, West Bengal - 721302, India*

Understanding and modulating the cross-stream motion of a surfactant-coated droplet in pressure driven flow has great implications in many practical applications. A combination of interfacial viscosity and Marangoni stress acting over a surfactant-coated droplet in pressure driven flow offers greater flexibility to modulate the cross-stream motion of it. Despite the intense theoretical and numerical research towards manipulating the surfactant-laden Newtonian droplets in Poiseuille flow, the experimental investigations are seldom explored. Herein, we report our study on understanding the influence of interfacial viscosity on the cross-stream motion of a surfactant-coated Newtonian droplet in both isothermal and non-isothermal Poiseuille flow from a theoretical as well as an experimental perspective. A theoretical model has been developed to understand the effect of interfacial viscosity on the lateral migration of a droplet under the assumptions of no shape deformation and negligible fluid inertia or thermal convection. Theoretical analysis is performed under two limiting conditions: (i) when the transport of surfactants is dominated by surface-diffusion and (ii) when the transportation of surfactants is dominated by surface-convection. Our theoretical analysis shows that both the dilatational as well as the shear surface viscosities suppress the lateral migration velocity of the droplet. Experiments have been performed to validate the theoretically predicted droplet trajectories and to understand the influence of channel confinement on the lateral migration of the droplet. It has been observed from the experiments that the droplet travels faster towards the centerline of the flow in a highly confined domain. The results presented in this study could provide new vistas in designing and analyzing various droplet-based microfluidic, biomedical and bio-microfluidic devices.

† E-mail address for correspondence: suman@mech.iitkgp.ernet.in

# 1. Introduction

In the present era, understanding the interfacial dynamics of droplets appears as an increasingly popular domain of research due to its diverse range of applications in different bio-microfluidic devices (Stone et al. 2004; Baroud et al. 2010; Seemann et al. 2012). Owing to the advancements in the droplet generation technologies, these devices are now making increasing use of droplets for performing tasks such as controlled delivery of drugs, encapsulation of biological cell, analytical detection etc.(Di Carlo *et al.* 2007; Huebner *et al.* 2008; Pethig 2013; Tao *et al.* 2015; Zhu &Fang 2013). The dynamics of droplets is also found to be important in the areas of biomolecules synthesis, mimicking the dynamics of vesicles and single cell analysis (Teh et al. 2008; Wyatt Shields IV et al. 2015; Casadevall i Solvas & DeMello 2011; Huebner et al. 2008). Knowledge of the motion of droplets in an imposed flow is required for developing an understanding of several naturally occurring processes like the cross-stream motion and positioning of erythrocytes in the microvasculature system (Fåhraeus 1929; Pries et al. 1996). Control over the motion of droplets and other suspended particles also has implications in flow cytometry and fractionalization of flow field ( Yang et al. 1999; Bonner et al. 1972).

There have been several theoretical and experimental studies on the dynamics of droplets in back ground pressure driven flow. Hetsroni & Haber (1970) performed an analytical study where they considered a spherical Newtonian droplet in an infinite Poiseuille flow. They have only studied the axial migration of the droplet and concluded that under the creeping flow limit, the migration of the droplet takes place towards the flow direction. However, several interesting phenomena can be observed when non-linear effects such as viscoelasticity, Marangoni stress, inertia and deformability are considered. One interesting fact is that the migration of an deformable droplet placed in a eccentric position takes place both in the direction of the flow as well as along the cross-stream wise direction (Griggs *et al.* 2007; Haber & Hetsroni 1971; Mandal *et al.* 2015; Mortazavi & Tryggvason 2000). For a clean droplet (free of surfactant), the lateral migration is solely influenced by the viscosity ratio of the droplet phase and suspending phase. In a related study, Chan & Leal (1979 ) had observed that the droplet moves away from the flow centerline for the values of viscosity ratio $\lambda$ (here $\lambda$ is the viscosity ratio between the droplet and carier phase) between 0.5 and 1. However, for all other values of $\lambda$, it is found that the motion of the droplet occurs towards the flow centerline. The existence of inertia (Karnis et al. 1966; Hur et al. 2011; Chen et al. 2014) and viscoelastic nature of the fluid (Chan & Leal 1979; Mukherjee & Sarkar 2013; Mukherjee & Sarkar 2014) are also found to exert significant influence on the cross-stream motion the of droplet.

It has been shown that the cross-stream motion of droplets in pressure driven flow can be controlled effectively by applying an external temperature gradient (Das et al. 2018). The presence of external temperature gradient leads to the alteration of temperature distribution on the surface of the droplet. This spatial variation in temperature gives rise to Marangoni stresses along the droplet's interface. The variation of temperature along the flow is found to exert a strong influence on the motion of droplets. From the study of Young et al. (1959), it has been observed that the solo presence of external temperature gradient can cause the cross-stream motion of the droplet. Following the work of Young *et al.*(1959) , there have been a number of studies which have focussed on understanding thermocapillary motion of droplets in a quiescent medium. The thermocapillary motion of droplets in an imposed poiseulle flow has been studied analytically by Raja Shekhar and co workers (Choudhuri & Raja Sekhar 2013; Sharanya & Raja Sekhar 2015). They have neglected the presence of surfactants and droplet deformation and have shown that under the creeping flow limit, the effect of external temperature gradient and inciepient flow can be linearly superimposed.

Surfactants are be frequently encountered in various droplet based microfluidic applications. They are either present naturally or are introduced as additives to enhance the stability of emulsions (Baret 2012). Along with lowering the interfacial tension, the uneven distribution of surfactants on the interface develops Marangoni stresses (Leal 2007). Both the thermal and the surfactant-induced Marangoni stresses have significant influence on the cross-stream motion of droplets (Das *et al.* 2017, 2017b). Therefore, a fundamental understanding of the influence of Marangoni stresses on the lateral migration of droplets is of paramount importance. A detailed study related to surfactant-induced cross-stream motion of the droplet is reported in the study by Hanna & Vlahovska (2010) and Pak *et al.*(2014). They have performed analytical studies to conclude that the marangoni stresses developed due to the uneven distribution of surfactants can develop a cross-stream motion of droplets even in the absence of deformation, viscoelasticity and other non-linear effects.

Interfacial viscosity refers to the resitance of the interface to deform under the application of stresses. Flumerfelt (1980) had characterized the interfacial viscosity of membranes and studied its influence on droplet deformation and orientation. In this study, Flummerfelt has provided a mathematical description of the extra interfacial viscous-stresses arising due to the interfacial viscosity. In the work of Ponce-Torres *et al.*(2017), the impact of surface viscosity on the breakup of a pendant droplet has been analyzed both theoretically and experimentally. They have reported the accumulation of surfactants in the resulting satellite droplet and have been able to explain it theoretically by incorporating the effect of interfacial viscosity. Recently Das & Chakraborty (2018) have performed an analytical and numerical study for investigating the influence of surface viscosity on the axial migration of a droplet in a non-isothermal pressure driven flow. However, the influence of interfacial viscosity on the cross-stream migration of surfactant-coated droplets in an unbounded non-isothermal Poiseuille flow is seldom explored.

In the our current study, we have developed a theoretical model to elucidate the role of interfacial viscosity on the cross-stream motion of a spherical Newtonian droplet in an unconfined non-isothermal pressure driven flow. For the theoretical analysis, an asymptotic approach is employed in order to tackle the high degree of non-linearity arising due to the transport of surfactants and the presence of interfacial viscosity. The analysis is performed under two limiting conditions. The first condition is that the transport of surfactant is dictated by surface-diffusion and the other is that the surfactant transport is dictated by the surface-convection mode of transport. Towards studying the effect of interfacial viscosity, two distinct parameters are identified namely: the dilatational Boussinesq number ($Bo_d$) and the shear Boussinesq number ($Bo_s$). It has been observed that the dilatational Boussinesq number exerts a stronger influence on droplet motion as compared to the shear Boussinesq number. The magnitude of cross-stream migration velocity of the droplet is found to decrease as both shear as well as dilatational Boussinesq numbers are increased. However, the magnitude of decrease in the lateral velocity of the droplet due to the former is far less as compared to that occurring due to the later. The nature of the decrease in the lateral migration velocity caused due to $Bo_s$ is found to be dependent on the surrounding temperature field. Therefore, it can be concluded that the existence of interfacial viscosity causes the reduction in the cross-stream migration velocity of the droplet and hence leads to a more naturalistic model for droplet migration in a non-isothermal flow. Furthermore, experiments are performed to validate the theoretical model as well as to analyze the trajectory of the droplet in tightly confined domiain. These experiments are performed for different degrees of channel confinement. The results obtained from the theoretical analysis match well with the experimental data when the confinement ratio is very low. Further, the effect of confinement on the cross-stream motion of the droplet is also shown through the experimental results.

## 2. Theoretical model

### 2.1. *System description*

In the present analysis, we have considered a system, where a spherical droplet of radius *a,* suspended in another fluid medium is experiencing combined presence of background pressure driven flow and axial temperature gradient. The fluids are considered to be Newtonian and incompressible in nature. The properties of fluids are density $\rho_i$, $\rho_e$; viscosity $\mu_i$, $\mu_e$ and thermal conductivity $k_i$, $k_e$. The subscripts '*i*' and '*e*' denote the droplet phase and ambient fluid phase respectively. The value of $\bar{e}$ denotes the distance between the droplet centroid and the flow centerline. The value of $\bar{e}$ is determined from $\bar{e} = \bar{x}_d - \bar{H}/2$. Here $\bar{x}_d$ refers to the distance of the droplet centroid from the wall of the microchannel (measured experimentally) and $\bar{H}$ refers to the width of the microchannel. The interface of the droplet is laden by bulk-insoluble surfactants having local concentration $\bar{\Gamma}$. The uniform concentration of surfactants ($\bar{\Gamma}_{eq}$) and the corresponding surface tension at the interface ($\bar{\sigma}_{eq}$) is disturbed due to the imposed fluid flow as well as the non-uniform temperature distribution at the interface ($\bar{T}_s$). The imposed flow alters the surfactant concentration via the convection mode of surfactant transport across the interface. The temperature field can alter the interfacial tension in two ways : (i) it directly affects the surface tension, and (ii) the thermally induced Marangoni stresses alter the surfactant transport across the interface that also causes a change in the local surface tension. The imposed Poiseuille flow is denoted by $\bar{\mathbf{V}}_\infty$, the temperature field is denoted by $\bar{T}_\infty$ and the temperature of the cold end is kept constant at $\bar{T}_c$. All the variables with an over bar denote the dimensional quantities and the ones without it represent the non-dimensional quantities. In the present study, the interfacial viscosity is assumed to remain constant across the interface. The interfacial viscosity is quantified by the values of $Bo_d = \mu_d/\mu_e a$ and $Bo_s = \mu_s/\mu_e a$ which are known as the dilatational and shear Boussinesq number respectively (Das & Chakraborty 2018). These are non-dimensional ways of expressing the dilatational and shear viscosities. The dilatational viscosity is the property of an interface due to which it resists expansion and compression. The shear Boussinesq number is physically understood as the resistance of the interface against angular deformation. Therefore, in the theoretical analysis, the effects of $Bo_d$ and $Bo_s$ are investigated on the lateral migration of a droplet in a plane-Poiseuille flow that is acted upon simultaneously by thermal and surfactant induced Marangoni stresses.

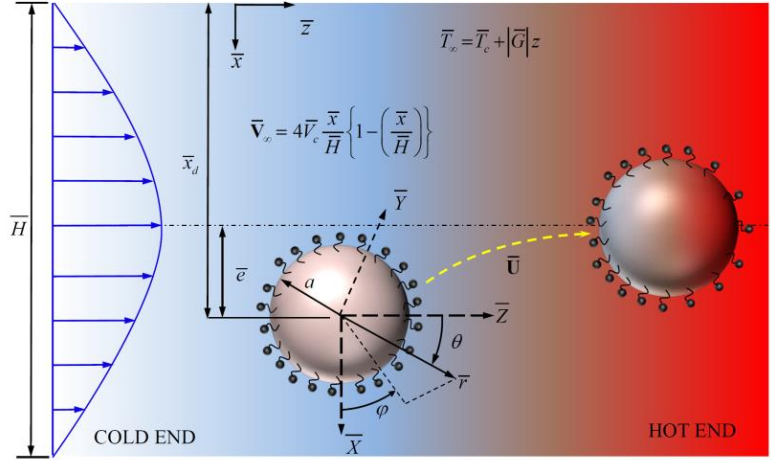

FIGURE 1. Schematic representation of a surfactant-coated droplet having radius $a$ suspended in a Plane-Poiseuille flow is shown. There is a linear change of temperature along the flow direction and $|\bar{G}|$ denotes the constant temperature gradient. A spherical co-ordinate system ($r$, $\theta$, $\varphi$) is attached to the center of the droplet. $\bar{e}$ denotes the eccentricity of the droplet and $\bar{H}$ denotes the width of the microchannel. $\bar{x}_d$ symbolizes the distance of the droplet centroid from the wall of the microchannel.

## 2.2. Assumptions

The important assumptions taken in the theoretical framework are: (i) The effects of fluid inertia are neglected and the pressure and viscous forces govern the flow problem. This implies that the Reynolds number ($Re = \rho \bar{V}_c a/\mu_e$) is very small in magnitude ($Re \ll 1$). Here $\bar{V}_c$ refers to the velocity at the centerline of the flow. (ii) The convection of thermal energy is considered to be negligible in comparison to diffusion, which results in the thermal Péclet number ($Pe_T = \bar{V}_c a/\alpha_e$) to be negligible. Here $\alpha_e$ refers to the thermal diffusivity of the suspending fluid. (iii) Natural convection is neglected as both the Grashof number ($Gr = g\gamma_e\rho_e^2 \Delta T a^3/\mu_e$) and the Rayleigh Number ($Ra = g\gamma_e\rho_e\Delta T a^3/\mu_e\alpha_e$) are very small ($Gr$, $Ra \ll 1$). Here, $\Delta T$ denotes the characteristic temperature difference. The volumetric expansion coefficient of the ambient fluid is denoted by $\gamma_e$. (iv) The shape of the droplet does not deviate from its spherical shape as the interfacial tension is much higher than viscous stresses. Hence, the capillary number $Ca = \mu_e \bar{V}_c / \bar{\sigma}_0$ can be assumed to be very small ($Ca \ll 1$). Here $\bar{\sigma}_0$ refers to the interfacial tension measured at the reference temperature $\bar{T}_0$. (v) we have also neglected the effect of wall confinement. (vi)The surfactants are bulk-insoluble and a linear relationship is assumed to hold between the surface tension and the surfactant concentration (Kim & Subramanian 1989)

$$\bar{\sigma} = \bar{\sigma}_0 - \beta(\bar{T}_s - \bar{T}_0) - R_g \bar{T}_0 \bar{\Gamma}, \qquad (1)$$

where $\bar{T}_s$ represent the temperature at the droplet interface and the ideal gas constant is denoted by $R_g$. $\beta$ is expressed as $\beta = -d\bar{\sigma}/d\bar{T}_s$. In our experimental study, we have taken a DI water droplet suspended in silicon oil having $\rho_e$ = 971 kg·m$^{-3}$, $\mu_e$ = 0.04855 Ns·m$^{-2}$, $\alpha_e$ = 7×10$^{-8}$ m$^{-2}$·s$^{-1}$ and $\gamma_e$ = 10$^{-4}$ (Huang & Liu 2009; Nallani & Subramanian 1993). The centerline velocity varies in the range of $O(10^{-4})$ - $O(10^{-5})$ ms$^{-1}$. For surfactant, we have chosen Triton X-100. The obtained non-dimensional parameters are : $Re \sim O(10^{-4})$, $Ca \sim O(10^{-4})$, $Pe_T \sim O(10^{-2})$, $Gr \sim O(10^{-6})$ and $Ra \sim O(10^{-2})$. The order of magnitude of non-dimensional numbers clearly justifies our assumptions.

2.3. *Governing equations and boundary conditions in non-dimensional format*

The non-dimensional format adopted in this study is:

$$\begin{aligned} r = \bar{r}/a, \mathbf{u} = \bar{\mathbf{u}}/\bar{V}_c, T = (\bar{T}-\bar{T}_o)/|\bar{G}|a, \Gamma = \bar{\Gamma}/\bar{\Gamma}_{eq}, \sigma = \bar{\sigma}/\bar{\sigma}_o, \\ p = \bar{p}/(\mu_e \bar{V}_c a) \, and \, \tau = \bar{\tau}/(\mu_e \bar{V}_c a). \end{aligned} \qquad (2)$$

Constant property ratios appearing are $\lambda = \mu_i/\mu_e$ which denotes the viscosity ratio of droplet phase and ambient fluid phase and $\delta = k_i/k_e$ which refers to ratio of thermal conductivity of the droplet phase and the ambient fluid phase. The important non-dimensional numbers are: (i)Surface Peclet Number $Pe_s = \bar{V}_c a/D_s$ ($D_s$ is the surface diffusivity of the surfactant) which signifies the relative strength of the convective mode of surfactant transport to the diffusive mode of surfactant transport along the interface. (ii)Surfactant Marangoni Number $Ma_\Gamma = \bar{\Gamma}_{eq} R_g \bar{T}_0 / \mu_e \bar{V}_c$ that gives the ratio of the strength of the surfactant-induced non-uniform interfacial tension driven Marangoni convection to the strength of the incipient flow. (iii)Thermal Marangoni Number $Ma_T = \beta|\bar{G}|a/\mu_e \bar{V}_c$ which is the ratio of the strength of the non-uniform temperature driven Marangoni flow and the imposed Poiseuille flow. (iv) Dilatational Boussinesq number $Bo_d = \mu_d/\mu_e a$ which signifies the relative strength of the interfacial dilatational viscous stress with respect to the bulk viscous stress. The shear Boussinesq number is defined as $Bo_s = \mu_s/\mu_e a$, which shows the relative strength of the interfacial shear stress to the bulk viscous stress. In order to theoretically investigate the effects of interfacial viscosity, the magnitude of the dilatational Boussinesq number $Bo_d$ is varied from 0.1 to 10 and the value of the shear Boussinesq number $Bo_s$ is varied from 0.01 to 10. From the experimental data reported in Ponce-Torres *et al.* 2017, it can be inferred that the value of $Bo_d$ and $Bo_s$ can vary from $O(10^{-2})$ to $O(10^3)$. However, we have restricted our results to the above-mentioned ranges in order to depict the most significant variations in the lateral migration velocity because of the presence of interfacial viscosity.

After employing the non-dimensional scheme given in equation(2), the following form of the governing differential equations and related boundary conditions are obtained:

The distribution of temperature is governed by

$$\left.\begin{aligned}\nabla^2 T_i &= 0, \\ \nabla^2 T_e &= 0,\end{aligned}\right\}, \tag{3}$$

subjected to following boundary conditions,

$$\left.\begin{aligned}&\text{as } r \to \infty, \ T_e = \zeta r \cos(\theta), \\ &T_i \text{ is bounded at } r=0, \\ &\text{at } r = 1, \ T_i = T_e, \\ &\text{at } r = 1, \ \delta\left(\nabla T_i \cdot \hat{n}\right) = \nabla T_e \cdot \hat{n}\end{aligned}\right\}. \tag{4}$$

Here the factor $\zeta$ denotes the direction in which temperature gradient is applied and $\hat{n}$ denotes the unit vector normal to the droplet surface. $\zeta$ can take binary values, 1 and -1. The value of $\zeta =1$ denotes the enhancement of temperature along the flow direction whereas $\zeta =-1$ denotes the reduction of temperature along the flow direction. The distribution of velocity field and pressure field is obtained by solving the following equations

$$\left.\begin{aligned}-\nabla p_i + \lambda \nabla^2 \mathbf{u}_i &= 0, \quad \nabla \cdot \mathbf{u}_i = 0, \\ -\nabla p_e + \nabla^2 \mathbf{u}_e &= 0, \quad \nabla \cdot \mathbf{u}_e = 0,\end{aligned}\right\}. \tag{5}$$

The related boundary condition are read as

$$\left.\begin{aligned}&\text{at } r \to \infty, \ (\mathbf{u}_e, p_e) = (\mathbf{V}_\infty - \mathbf{U}, \ p_\infty), \\ &\mathbf{u}_i \text{ is bounded at } r=0, \\ &\text{at } r=1, \ \mathbf{u}_i \cdot \mathbf{e}_r = \mathbf{u}_e \cdot \mathbf{e}_r = 0, \\ &\text{at } r=1, \ \mathbf{u}_i = \mathbf{u}_e, \\ &\text{at } r=1, \ (\boldsymbol{\tau}_e \cdot \mathbf{e}_r - \boldsymbol{\tau}_i \cdot \mathbf{e}_r) \cdot (\mathbf{I} - \mathbf{e}_r \mathbf{e}_r) = Ma_\Gamma (\nabla_s \Gamma) \cdot \mathbf{e}_\theta + Ma_T (\nabla_s T_s) \cdot \mathbf{e}_\theta - (\nabla_s \cdot \boldsymbol{\tau}_s) \cdot \mathbf{e}_\theta.\end{aligned}\right\}. \tag{6}$$

Based on the Boussinesq-Scriven constitutive law for Newtonian fluids, the surface excess viscous stress can be read as

$$\bar{\boldsymbol{\tau}}_s = 2\mu_s \bar{\mathbf{D}}_s + (\mu_d - \mu_s)\{\mathbf{I}_t : \bar{\mathbf{D}}_s\}\mathbf{I}_t, \tag{7}$$

where the dilatational and shear viscosities at the droplet interface are denoted by $\mu_d$ and $\mu_s$ respectively. The rate of deformation tensor is symbolizes by $\bar{\mathbf{D}}_s$ and read as

$$\bar{\mathbf{D}}_s = \frac{1}{2}\left\{ \bar{\nabla}_s \bar{\mathbf{u}} \bullet \mathbf{I}_t + \mathbf{I}_t \bullet \left(\bar{\nabla}_s \bar{\mathbf{u}}\right)^T \right\} \qquad (8)$$

After following the non-dimensionalising scheme given in equation (2), the following non-dimensional form is obtained

$$\nabla_s \cdot \boldsymbol{\tau}_s = \left(\mathbf{I}_t \cdot \nabla\right) \cdot \left\{ \left(Bo_d - Bo_s\right)\left(\mathbf{I}_t : \nabla\right)\mathbf{I}_t + 2Bo_s\left(\mathbf{I}_t \cdot \mathbf{D}_s \cdot \mathbf{I}_t\right)\right\}. \qquad (9)$$

Equation (9) represents the general expression for surface excess viscous stresses, by assuming that the values of $Bo_d$ and $Bo_s$ are constant, the following equation is obtained (Flumerfelt 1980)

$$\nabla_s \cdot \boldsymbol{\tau}_s = 2Bo_s\left\{\left(\mathbf{I}_t \cdot \nabla\right) \cdot \left(\mathbf{I}_t \cdot \mathbf{D}_s \cdot \mathbf{I}_t\right)\right\} + \left(Bo_d - Bo_s\right) \times \left\{\mathbf{I}_t \cdot \nabla\left(\mathbf{I}_t : \nabla \mathbf{u}\right) + 2H\left(\mathbf{I}_t : \nabla \mathbf{u}\right)\right\}, \qquad (10)$$

here $H$ is the mean curvature and $H = -1$ for a spherical droplet.

The surfactant concentration at the interface is governed by the surfactant transport equation, which can be expressed in its non-dimensional form as

$$Pe_s \nabla_s \cdot \left(\mathbf{u}_s \Gamma\right) = \nabla_s^2 \Gamma. \qquad (11)$$

Here $\mathbf{u}_s$ refers to the interfacial fluid velocity of the droplet. In addition to that the surfactant concentration must also satisfy the conservation of mass which is given by

$$\int_{\varphi=0}^{2\pi}\int_{\theta=0}^{\pi} \Gamma(\theta,\varphi)\sin\theta\, d\theta\, d\varphi = 4\pi \qquad (12)$$

## 3. Asymptotic solution

In this section, a brief outline of the solution methodology is presented along with the important analytical results.

Since the temperature field satisfies the Laplace equation, therefore it can be represented as a linear combination of the spherical harmonics (Choudhuri & Raja Sekhar 2013). In a similar spirit the surfactant concentration is also expanded in terms of spherical harmonics (Haber & Hetsroni 1972; Pak et al. 2014). The velocity field in the inner side of the droplet satisfies the stokes and continuity equations, hence it can be represented by the growing spherical harmonics (Hetsroni & Haber 1970). Similarly, the velocity field external to the droplet is expressed as a superposition of the far-field velocity field and the decaying spherical harmonics.

### 3.1. Solution for $Pe_s \ll 1$

From equations (3) and (4), it is evident that the temperature field in and outside of the droplet does not depend on the velocity field and the interfacial surfactant distribution. Therefore, they are solved independently without any regard to the fluid flow and surfactant transport equations. The temperature field is obtained as follows

$$\left. \begin{array}{l} T_i = \zeta \left( \dfrac{3}{\delta+2} \right) r P_{1,0} (\cos\theta), \\ T_e = \zeta \left[ r + \left( \dfrac{1-\delta}{2+\delta} \right) \dfrac{1}{r^2} \right] P_{1,0} (\cos\theta). \end{array} \right\} \quad (13)$$

A regular perturbation analysis is employed. Any dependent variable $\Phi$ is represented by(Pak et al. 2014):

$$\Phi = \Phi^{(0)} + Pe_s \Phi^{(Pe_s)} + Pe_s^2 \Phi^{(Pe_s^2)} + O(Pe_s^3). \quad (14)$$

The approach used in this study has been successfully applied by Das *et al.* (2018a) to obtain the lateral migration velocity of a droplet suspended in a non-isothermal plane poiseuille flow. However, in the present study, the focus is on studying the influence of the extra stress terms arising due to interfacial viscosity. It is given by equation(10). The RHS of equation (10) is evaluated by expanding the interfacial velocity as given by Lamb's general solution(Lamb 1895).

In the low Peclet regime, the surfactant transport equation at any order of perturbation does not depend the velocity field at that order, hence it is solved before the solving the flow field equations. At the leading order the surfactant concentration is obtained as

$$\Gamma^{(0)} = 1. \quad (15)$$

In the absence of buoyancy forces, the net force on the droplet is a result of the hydrodynamic forces due to pressure and viscous stresses. It is represented by $\boldsymbol{F}_H$. The leading order velocity field is solved, after obtaining the velocity field, the migration velocity of the droplet at the leading order is obtained using the force-free condition:

$$\boldsymbol{F}_H^{(0)} = 0 \Rightarrow -4\pi \nabla (r^3 p_{-2}^{(0)}) = 0. \quad (16)$$

Here $\boldsymbol{F}_H^{(0)}$ refers to the leading order hydrodynamic force that acts on the droplet.

The leading order migration velocity of the droplet is obtained after solving equation (16) and is expressed as

$$U_z^{(0)} = \left[ \begin{array}{c} \underbrace{\left( \dfrac{4\left(2eH - 2e^2 - \lambda + 3eH\lambda - 3e^2\lambda\right)}{(2+3\lambda)H^2} \right)}_{\text{Effect of Imposed Flow}} + \underbrace{\left( \dfrac{2Ma_T \zeta H^2}{(2+3\lambda)(\delta+2)H^2} \right)}_{\text{Effect of Temperature Gradient}} \\ \underbrace{-\left( \dfrac{4}{3} \dfrac{Bo_d \left(3Ma_T \zeta H^2 + 8 + 4\delta\right)}{(2+3\lambda - 2Bo_d)(\delta+2)H^2(2+3\lambda)} \right)}_{\text{Effect of Interfacial Viscosity}} \end{array} \right],$$

$$U_x^{(0)} = U_y^{(0)} = 0,$$
(17)

here $U_z^{(0)}$ represents the leading order axial component of the droplet migration velocity, $U_x^{(0)}$ is the leading order lateral migration velocity. It can be concluded from equation (17) that the shear surface viscosity does not influence the leading order axial migration of the droplet. The $O(Pe_s)$ surfactant concentration is achieved by substituting the leading order surface velocity into the $O(Pe_s)$ surfactant transport equation. The $O(Pe_s)$ surfactant concentration is obtained as

$$\Gamma^{(Pe_s)} = \Gamma_{1,0}^{(Pe_s)} P_{1,0}(\cos\theta) + \Gamma_{2,1}^{(Pe_s)} P_{2,1}(\cos\theta)\cos\phi + \hat{\Gamma}_{2,2}^{(Pe_s)} P_{2,2}(\cos\theta)\sin 2\phi \\ + \Gamma_{3,0}^{(Pe_s)} P_{3,0}(\cos\theta) + \Gamma_{3,2}^{(Pe_s)} P_{3,2}(\cos\theta)\cos(2\phi),$$
(18)

where the constant coefficients appearing in equation (18) are mentioned in the supplementary material.

The $O(Pe_s)$ velocity field is obtained by solving the $O(Pe_s)$ governing equations and related boundary conditions. The migration velocity is achieved by solving the force-free condition which is given as follows

$$\mathbf{F}_H^{(Pe_s)} = -4\pi \nabla \left( r^3 p_{-2}^{(Pe_s)} \right) = \mathbf{0}.$$
(19)

Here $\mathbf{F}_H^{(Pe_s)}$ refers to the hydrodynamic force at $O(Pe_s)$.

The $O(Pe_s)$ droplet migration velocity is obtained as follows

$$U_z^{(Pe_s)} = -\dfrac{2}{9} \left( \dfrac{Ma_\Gamma \left(9Ma_T \zeta H^2 + 24 + 12\delta\right)}{(2 + 2Bo_d + 3\lambda)(6\lambda + 3\delta\lambda + 4 + 4Bo_d + 2\delta + 2Bo_d\delta)H^2} \right),$$

$$U_x^{(Pe_s)} = U_y^{(Pe_s)} = 0.$$
(20)

Finally the $O(Pe_s^2)$ migration velocity is obtained by following a similar procedure and the $O(Pe_s^2)$ migration velocity is obtained as

$$\left. \begin{aligned} U_z^{(Pe_s^2)} &= \frac{2}{9}\left( \frac{Ma_\Gamma^2\left(24\lambda+12\lambda\delta-9Ma_T\zeta H^2\lambda\right)}{\lambda(3\lambda-2Bo_d+2)\left((3\lambda+2)^2(2+\delta)-4Bo_d(-Bo_d+3\lambda+2)(2+\delta)\right)H^2} \right), \\ U_x^{(Pe_s^2)} &= \frac{\sum_{i=0}^{7}\left(\sum_{j=0}^{7} f_{i,j} Bo_d^{\,j}\right)Bo_s^{\,i}}{\Phi(Bo_d, Bo_s, \lambda, \delta)}, \\ U_y^{(Pe_s^2)} &= 0. \end{aligned} \right\} \quad (21)$$

The constants appearing in equation (21) are mentioned in the supplementary material.

The $O(Pe_s^2)$ surfactant concentration is obtained as

$$\left. \begin{aligned} \Gamma^{(Pe_s^2)} &= \Gamma_{1,0}^{(Pe_s^2)} P_{1,0}(\cos\theta) + \Gamma_{1,1}^{(Pe_s^2)} P_{1,1}(\cos\theta)\cos(\phi) + \Gamma_{2,0}^{(Pe_s^2)} P_{2,0}(\cos\theta) + \\ &\quad \Gamma_{2,1}^{(Pe_s^2)} P_{2,1}(\cos\theta)\cos(\phi) + \Gamma_{2,2}^{(Pe_s^2)} P_{2,2}(\cos\theta)\cos(2\phi) + \hat{\Gamma}_{2,1}^{(Pe_s^2)} P_{2,1}(\cos\theta)\sin(\phi) \\ &\quad + \Gamma_{3,0}^{(Pe_s^2)} P_{3,0}(\cos\theta). \end{aligned} \right\} \quad (22)$$

The constant coefficients appearing in equation (22) are not mentioned for the sake of brevity. The final form of the lateral migration velocity can be read as

$$U_x = U_x^{(0)} + Pe_s U_x^{(Pe_s)} + Pe_s^2 U_x^{(Pe_s^2)}. \tag{23}$$

The temporal variation of transverse position of the droplet is obtained by solving the following linear differential equation

$$\frac{dx}{dt} = U_x. \tag{24}$$

The final expression of is represented in the following form,

$$x = \frac{H}{2} + e^{at}\left(1 - \frac{H}{2}\right), \tag{25}$$

where

$$a = \frac{\sum_{i=0}^{4}\sum_{j=0}^{4-i} g_{i,j} Bo_s{}^j Bo_d{}^i}{d}. \qquad (26)$$

The constants appearing in equation (26) are mentioned in the supplementary material.

### 3.2. *Solution for $Pe_s \gg 1$*

Since the governing equations and related boundary conditions for the thermal problem are independent of the surface Peclet number, therefore the temperature distribution for the high Peclet regime is also given by equation (13). Here the surfactant Marangoni number is assumed to be very large in magnitude, i.e. $Ma_\Gamma \gg 1$ and $Ma_\Gamma^{-1}$ is taken to be the perturbation parameter for expanding the dependent variables in the problem (Hanna & Vlahovska 2010). Therefore, all the dependent variables other than the surfactant concentration are expressed in the following form

$$\chi = \chi^{(0)} + Ma_\Gamma^{-1} \chi^{(Ma_\Gamma^{-1})} + O(Ma_\Gamma^{-2}). \qquad (27)$$

The surfactant concentration is expressed as

$$\Gamma = 1 + Ma_\Gamma^{-1} \Gamma^{(0)} + Ma_\Gamma^{-2} \Gamma^{(Ma_\Gamma^{-1})} + O(Ma_\Gamma^{-3}). \qquad (28)$$

Owing to the nature of the surfactant transport equation at each order of perturbation, the surfactant-transport equation needs to be solved together along with the governing equations and boundary conditions for the flow-field. After solving for the flow field, we have utilized the force free condition on the droplet for obtaining the migration velocity of the droplet. The leading order migration velocity of the droplet is read as

$$\left. \begin{array}{l} U_z^{(0)} = \dfrac{2}{9}\left(\dfrac{-6+18eH-18e^2}{H^2}\right), \\ U_x^{(0)} = U_y^{(0)} = 0, \end{array} \right\} \qquad (29)$$

The surfactant concentration in leading order can be expressed is in the following form,

$$\left. \begin{array}{l} \Gamma^{(0)} = \Gamma_{1,0}{}^{(0)} P_{1,0}(\cos\theta) + \Gamma_{2,1}{}^{(0)} P_{2,1}(\cos\theta)\cos\phi + \hat{\Gamma}_{2,2}{}^{(0)} P_{2,2}(\cos\theta)\sin(2\phi), \\ + \Gamma_{3,0}{}^{(0)} P_{3,0}(\cos\theta) + \Gamma_{3,2}{}^{(0)} P_{3,2}(\cos\theta)\cos(2\phi), \end{array} \right\} \qquad (30)$$

where

$$\left.\begin{aligned}
\Gamma_{1,0}^{(0)} &= -\frac{8+4\delta+3Ma_T\zeta H^2}{H^2(\delta+2)}, \quad \Gamma_{2,1}^{(0)} = \frac{10}{3}\frac{H-2e}{H^2}, \\
\hat{\Gamma}_{2,2}^{(0)} &= \frac{40}{9}\frac{Bo_s}{H^2(\lambda+4)}, \quad \Gamma_{3,0}^{(0)} = \frac{7}{3H^2}, \quad \Gamma_{3,2}^{(0)} = -\frac{7}{18H^2}.
\end{aligned}\right\} \quad (31)$$

The $O(Ma_\Gamma^{-1})$ droplet migration velocity is obtained as

$$\left.\begin{aligned}
U_z^{(Ma_\Gamma^{-1})} &= 0, \\
U_x^{(Ma_\Gamma^{-1})} &= \frac{-2}{3}\left(\frac{(4(3\lambda+2)(\delta+2)+9H^2\zeta Ma_T(\lambda+4))(H-2e)}{(\lambda+4)(\delta+2)H^4}\right), \\
U_y^{(Ma_\Gamma^{-1})} &= 0.
\end{aligned}\right\} \quad (32)$$

The $O(Ma_\Gamma^{-1})$ surfactant concentration is represented as

$$\left.\begin{aligned}
\Gamma^{(Ma_\Gamma^{-1})} = &\Gamma_{1,1}^{(Ma_\Gamma^{-1})}P_{1,1}(\cos\theta)\cos\phi + \Gamma_{2,0}^{(Ma_\Gamma^{-1})}P_{2,0}(\cos\theta) + \Gamma_{2,2}^{(Ma_\Gamma^{-1})}P_{2,2}(\cos\theta)\cos 2\phi + \\
&\Gamma_{3,1}^{(Ma_\Gamma^{-1})}P_{3,1}(\cos\theta)\cos\phi + \Gamma_{3,3}^{(Ma_\Gamma^{-1})}P_{3,3}(\cos\theta)\cos 3\phi k.
\end{aligned}\right\} \quad (33)$$

The constant coefficients appearing in equation (33) are mentioned in the supplementary material.

## 4. Experimental setup and methodology

In this section, we discuss on the fabrication of PDMS microfluidic device, experimental setup and the methodology adopted to realize the cross-stream motion of the surfactant-coated droplet in non-isothermal pressure driven flow. All the experiments were performed in a controlled ambient condition of 27°C and 50 % humidity. All the fluids (50 cSt silicon oil and DI water) and chemicals (Triton X-100, nonionic surfactant) used were of analytical grade and used as received.

### 4.1. *Fabrication of micro fluidic device.*

The master mold of the microfluidic device is prepared by conventional photolithography process and the same is replicated by standard soft-lithography process to get polydimethylsiloxane (PDMS) device(Dey et al. 2015). Briefly, a borosilicate glass substrate is cleaned by using piranha solution ($H_2SO_4$:$H_2O_2$ in the ratio 1:1) to remove the organic contaminants followed by thorough rinsing with the deionized water to remove the residual dirt. The substrate is then spin coated using negative photo resist SU8 2150 (Micro Chem Corp, USA) at 3000 rpm for 25s to get a desired thickness of about 150-200 micrometer. Then, the substrate is subjected to prebaking at 65 °C for 7 minutes followed by 95 °C for 40 minutes over a hot-plate to evaporate the solvent. Further, the substrate is exposed to UV light (~ 365 nm, OAI 200 Mask aligner) through a chrome-printed photomask for 25 s followed by post-baking at 65 °C for 5 min and 95 °C for 15 min in a hot-air oven. Finally, the microchannel pattern mold is developed using SU8 developer solution. To obtain PDMS microfluidic device, standard soft-lithography approach is adopted, briefly, the components of silicone elastomer (Sylgard 184, Dow Corning, USA) base to curing agent were thoroughly mixed in the ratio 10:1 by weight. The mixture is then degassed by using a vacuum desiccators for removing the trapped air bubbles. Then it is transferred over the master mold and treated at 95 °C for 3 hours in a hot-air oven for curing. After solidification, the PDMS pattern is carefully peeled off from the master mold. The PDMS pattern is then bonded to a glass slide by oxygen-plasma bonding (Dey et al. 2015)

### 4.2. *Experimental setup and methodology*

To experimentally realize the cross-stream motion of surfactant-coated droplet in non-isothermal plane-Poiseuille flow as predicted by the theoretical analysis, an experimental setup is designed and implemented as depicted in the figure 2. The setup consists of a PDMS based microfluidic device for droplet generation, an inverted fluorescence microscope (Olympus IX71) coupled with the high-speed camera (Phantom V641) for recording the droplet migration trajectories, a strip heater (tungsten) connected with a voltage source meter (KEITHLEY- 2410) to generate temperature gradient and a temperature measurement system (T-type thermocouple and 3A972A Agilent LXI data acquisition system) to measure the surface temperature.

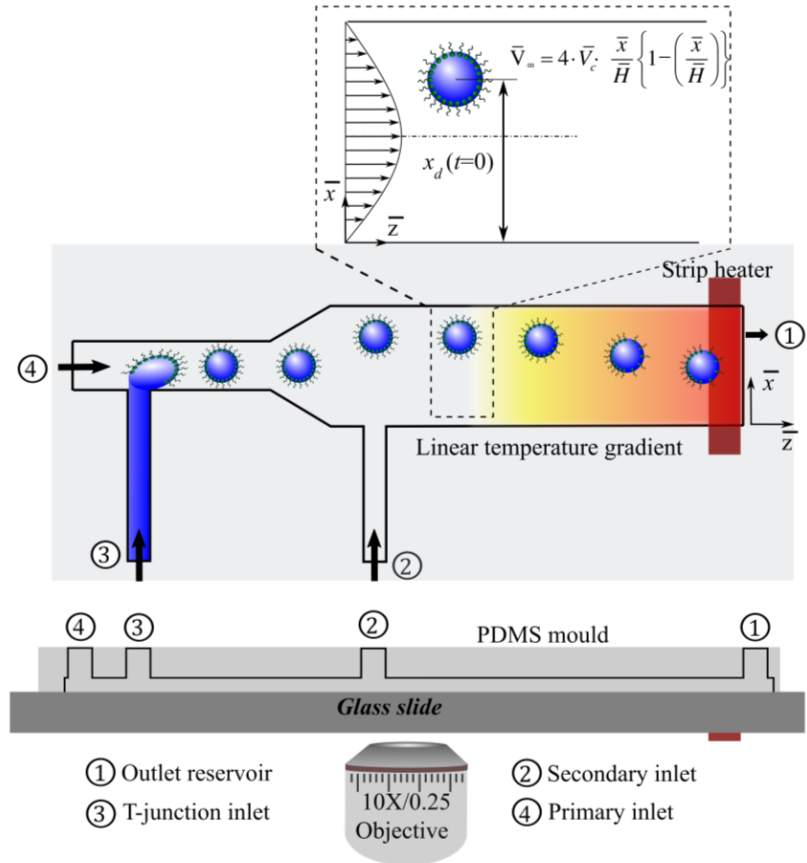

FIGURE 2. Schematic representation of the experimental setup developed for the present study. The top portion is the top view of the setup and bottom portion is the side view of the microfluidic device and setup.

The PDMS-glass microfluidic device comprises three inlet and one outlet port. Inlet ports are connected to the syringe pumps (Harvard Apparatus PHD 2000: 0-100 ml/min) whereas the outlet port is connected the reservoir through identical Tygon tubes. Inflow of carrier fluid (silicon oil) is created through the primary inlet and the dispersed fluid (a solution of DI water and Triton X-100 having concentration 400 ppm) is pumped through the T-junction inlet. Due to the interplay between the interfacial viscosity and the interfacial tension of continuous fluid and the dispersed fluid at the T-junction, a micrometer droplet is generated as shown in the figure 2. As the droplet moves to the wider cross section channel, the droplet is set offset by the inflow of silicon oil through secondary inlet. A nearly linear temperature field is generated along the direction of flow by heating a strip heater by applying the desired voltage. A series of grooves were made in the PDMS device near to wider cross section microchannel, and the thermocouples were inserted into the grooves to measure the temperature. Once the desired temperature gradient attains the steady state, the droplet migration trajectories were captured using the high-speed camera at 200 fps. Experiments were conducted to explore the influence of various confinement ratios.

## 5. Results and discussions

In this section, first we have made a comparison between the theoretical prediction considering the effect of interfacial viscosity and the experimental result on the trajectory of droplet centroid in a weakly confined domain. Next, we discuss the influence of interfacial viscosity on the lateral migration velocity of a droplet in both isothermal as well as non-isothermal Poiseuille flow. Here, we also experimentally show the effect of channel confinement on the trajectory of the droplet under both isothermal and non-isothermal conditions.

### 5.1. *Comparison between analytical and experimental result*

Figure 3(a) shows a comparison between our experimental observation and theoretical predictions (both with and without considering the effect of interfacial viscosity) on the temporal alteration of transverse position of the droplet centroid subjected to background pressure driven flow and axial temperature gradient.

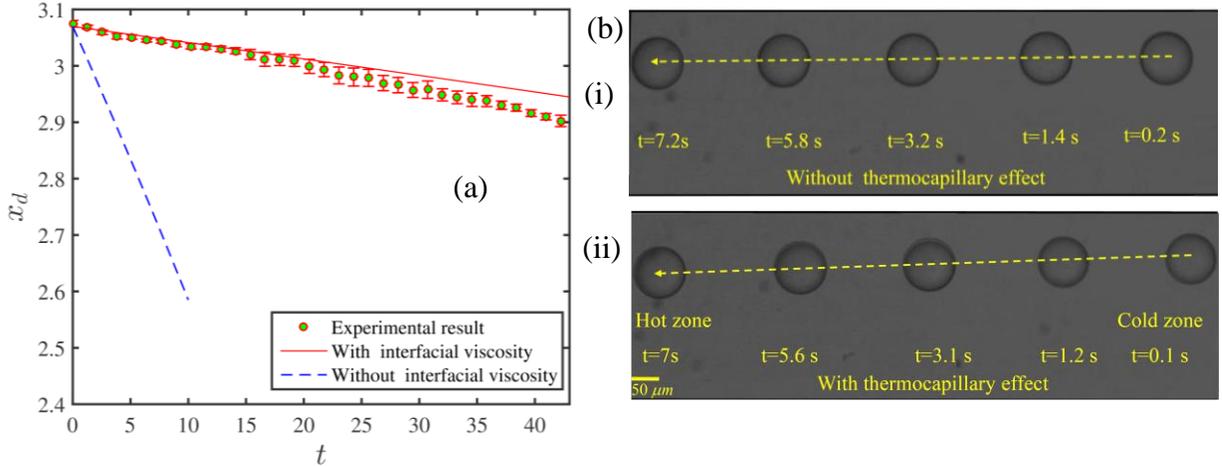

FIGURE 3. (a) Comparison of analytical and experimental result on transient variation of the trajectory of droplet centroid under axial temperature gradient at $Wc = 0.25$, (b) experimental images of the droplet at different time frames at $Wc=0.35$. Others parameters are $Ma_T = 5.5$, $Ma_\Gamma = 3.78$, $Bo_d = 3.28$, $Bo_s = 0.30$ and $Re \sim O(10^{-4})$.

The theoretical trajectory of the droplet without considering the effect of interfacial viscosity is obtained by equating $Bo_d = Bo_s = 0$. From the comparison, it is clearly seen that the present theoretical model considering the effect of interfacial viscosity provides better matching with the experimental results and offers a much greater approximation to the naturally occurring lateral migration of a droplet laden with bulk-insoluble surfactants and exposed to a non-isothermal Poiseuille flow. The slight deviation of the experimental result from the analytical one (with surface viscosity) is due to the presence of wall in the experimental analysis. In the next section, we have discussed about the impact of channel confinement on the trajectory of the droplet. Figures 3b(i) and 3b(ii) show the images of the droplet taken at different time intervals. Figure 3b(i) shows the droplet's images in an isothermal flow and figure 3b(ii) depicts the same when temperature gradient acts along the direction of the flow. On comparing the two figures, it is evident that the cross-stream motion of the droplet is more prominent in the second case, i.e. in presence of an applied temperature gradient.

### 5.2. *Impact of channel confinement on the trajectory of the droplet*

Here, we discuss the effect of the bounding walls on the cross-stream motion of the droplet. In confined domain, the analytical result is unable to predict the trajectory of the droplet accurately. Thus for doing this analysis, we have performed experiment. The effect of the bounding walls is quantified with the help of the domain confinement ratio ($Wc$), which is defined as the ratio of the droplet size to the height of the channel, i.e. $Wc = 2a/\bar{H}$.

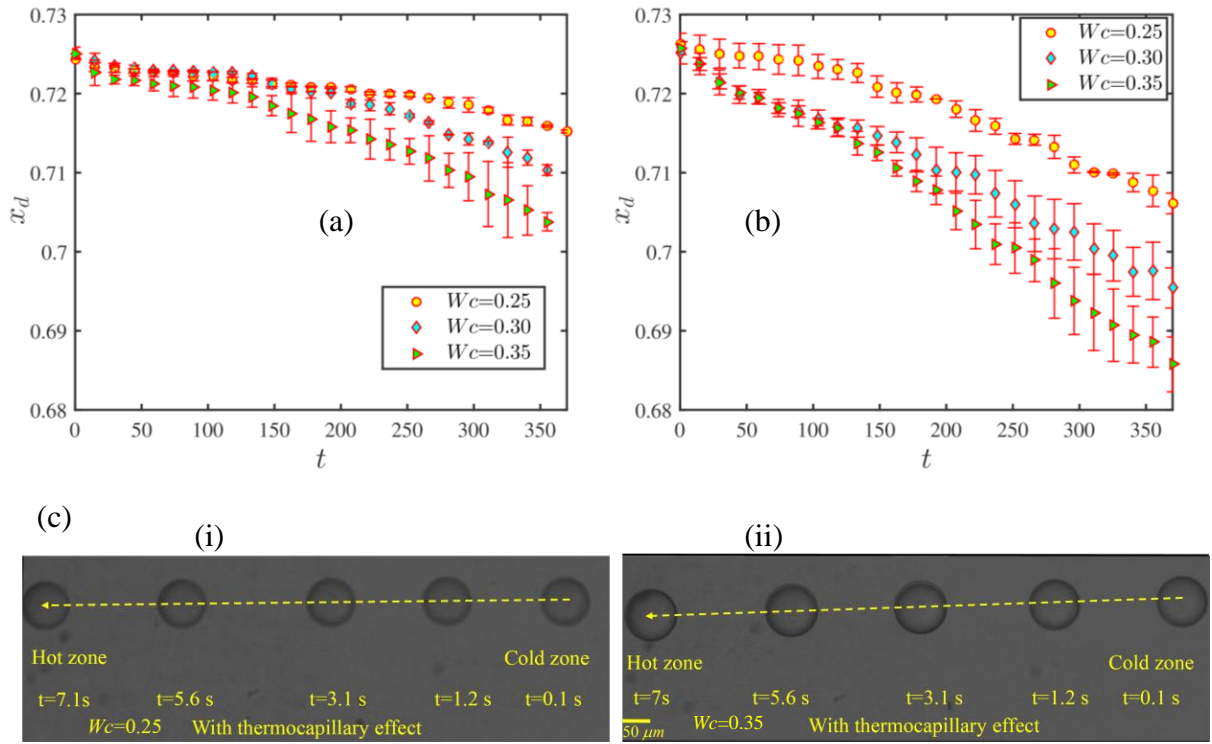

Figure 4(a) Temporal evolution of the droplet centroid in an isothermal Plane Poiseuille Flow. (b) Temporal evolution of droplet centroid in a non-isothermal flow with $Ma_T = 5.5$. (c) Snapshots a single droplet taken at different time intervals when (i) $Wc = 0.25$ and (ii) $Wc = 0.35$. The values of the other parameters are taken as $Ma_T = 3.78$, $Bo_d = 3.28$, $Bo_s = 0.30$, $Wc = 0.3$ and $Re \sim O(10^{-4})$.

The domain confinement ratio is varied by changing the droplet size by suitably adjusting the flow rate of the carrier fluid phase as explained in the study of Santra *et al.* (2018). Towards studying the influence of the domain confinement on the cross-stream motion of the droplet, the temporal variation of the experimentally measured transverse position of the droplet is plotted for $Wc$ equal to 0.25, 0.30 and 0.35 corresponding to $Q_d/Q_c$ equal to 0.17, 0.22 and 0.33 respectively. In order to understand the confluence of domain confinement and thermally induced Marangoni stresses on the droplet in confined domain, two plots are considered: i) the variation of the transverse position of the droplet centroid in an isothermal flow as shown in figure 4(a) and ii) the variation of the transverse position of the droplet centroid in a non-isothermal flow as depicted in figure 4(b). From these two figures it is noted that the migration of droplet towards channel centerline takes place very rapidly in highly confined domain i.e. a bigger droplet is found to move faster towards the centerline of the flow than a smaller droplet. Again, by comparing figures 4(a) and 4(b) we can also note that due to the presence of the axial temperature gradient, the lateral migration rate of the droplet is further enhanced. Figures 4c(i) and 4c(ii) show the snapshots of the droplet taken at different time intervals for a domain confinement ratio of 0.25 and 0.35 respectively, where the lateral migration of the droplet is more prominent for the latter case.

For a similar degree of surfactant transport along the interface of the droplet, the magnitude in the difference between the fluid flow velocities in the top and bottom hemispheres is higher for a greater droplet size. Therefore, a larger droplet exhibits a greater asymmetry in the distribution of surfactant along the interface and also results in a more asymmetrical temperature distribution along the interface. This results in a higher value of the Marangoni stresses, therefore resulting in a greater force that moves the droplet towards the flow centerline. Furthermore, the hydrodynamic lift force exerted by the channel wall also causes the faster migration of the droplet towards the centerline and its magnitude enhances with increase in the degree of confinement. Therefore, for a similar value of the axial temperature gradient, the cross-stream migration velocity is higher for a droplet having larger size. It is now known that the presence of an external temperature gradient results in a non-uniform distribution of the temperature along the interface of the droplet. A larger difference in the fluid flow velocities in the two hemispheres results in a larger asymmetry in the temperature distribution along the interface, which leads to a greater value of the thermally induced Marangoni stress along the interface of the droplet. Therefore, the increase in the lateral migration velocity of the droplet due to externally applied temperature gradient is much higher for a larger droplet as compared to a smaller droplet.

### 5.3. *Effect of interfacial viscosity*

#### 5.3.1. *Isothermal Poiseuille flow*

Figure 5 depicts the alteration of the cross-stream migration velocity of the droplet ($U_x$) with the viscosity ratio of the system ($\lambda$) for different dilatational Boussinesq number ($Bo_d$).

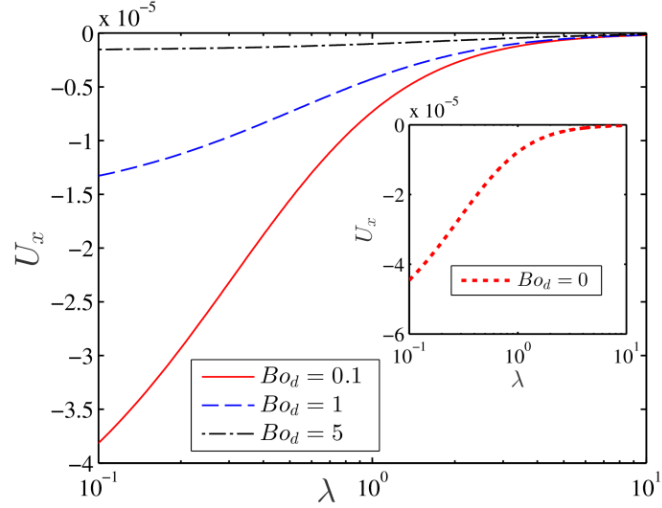

FIGURE 5. The variation of $U_x$ is plotted with $\lambda$ for various values of $Bo_d$ when $Ma_T = 0$, $e = 1$, $Pe_s = 0.1$, $Bo_s = 0$, $H = 10$, $Ma_\Gamma = 0.1$ and $\delta = 0.1$.

For the present analysis, four different magnitudes of $Bo_d$ have been considered. It can be observed that as the value of $Bo_d$ is increased, the magnitude of $U_x$ decreases. In the inset of figure 5, the variation of the cross-stream migration velocity is plotted where the effect of $Bo_d$ is neglected. It can be clearly observed that the magnitude of the cross-stream migration velocity is greater for all values of the viscosity ratio when the effect of $Bo_d$ is neglected. Therefore, the presence of dilatational interfacial viscosity actually suppresses the cross-stream migration velocity of the droplet and the magnitude of this suppression increases on increasing the value of the dilatational Boussinesq number. Figure 6 shows the alteration of $U_x$ with $\lambda$ for different values of the shear Boussinesq number ($Bo_s$). The magnitude of $U_x$ decreases on increasing the values of $Bo_s$. However, the magnitude of this decrease is much smaller in comparison to the decrease caused due to the dilatational Boussinesq number.

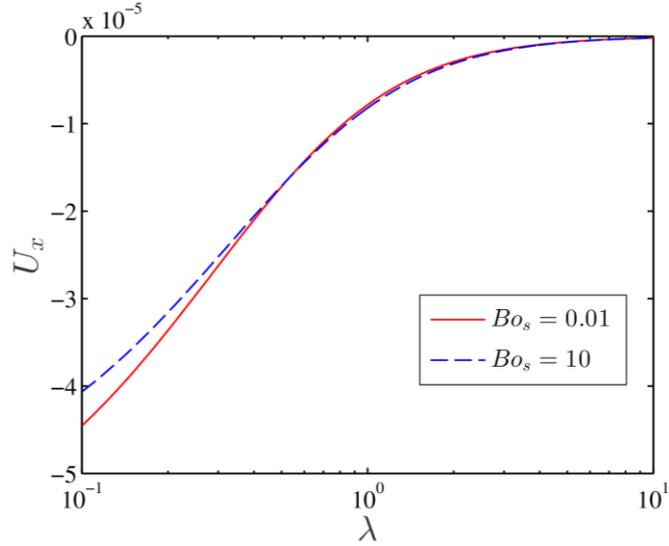

FIGURE 6. The variation of $U_x$ is plotted with $\lambda$ for various values of $Bo_s$ when $Ma_T = 0$, $e = 1$, $Pe_s = 0.1$, $Bo_d = 0$, $H = 10$, $Ma_\Gamma = 0.1$ and $\delta = 0.1$.

The physical reason behind the reduction of lateral migration velocity of the droplet due to interfacial viscosity can be identified, if we carefully observe the distribution of surfactant concentration at the droplet interface. Towards this, the contour plots for the surfactant concentration at the interface are plotted in figure 7. Figure 7(a) depicts the interfacial surfactant distribution for the case when $Bo_d = 0$ and figure 7(b) shows the interfacial surfactant distribution when $Bo_d = 10$. A closer look into the figures 7(a) and 7(b) reveals that the concentration of surfactant is maximum near the north-west pole and minimum near the north-east pole of the droplet. Therefore, the surfactant distribution for an eccentrically placed droplet is asymmetric both about the transverse as well as about the axial planes. The non-uniformity of surfactant distribution about the transverse plane results in retardation in the axial migration velocity of the droplet. However, the non-uniformity about the axial plane results in the lateral migration of the droplet. Therefore, in the subsequent discussions, more attention will be paid to the non-uniformity in the surfactant concentration about the axial plane. Owing to the closer proximity of the north-pole of the droplet to the centerline of the flow (see Figure 1), it is exposed to a higher imposed velocity than the south pole. This creates an asymmetry in the surfactant concentration, which in turn creates an asymmetric surface tension distribution at the droplet interface. This develops a hydrodynamic force in the cross-streamwise direction that causes the lateral migration of the droplet.

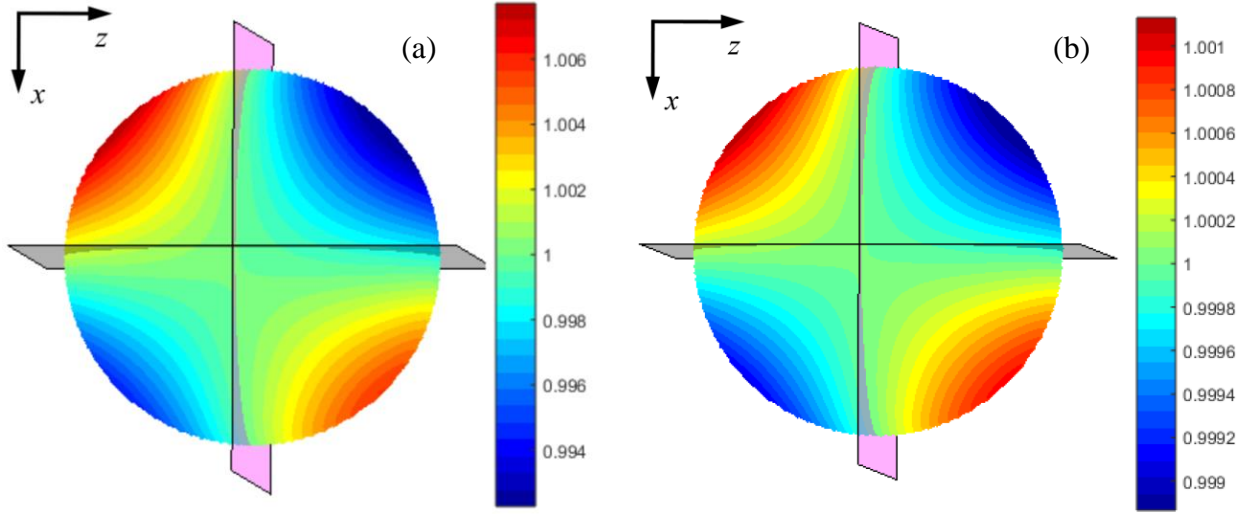

FIGURE 7. Surface plot of the surfactant concentration at the droplet interface for (a) $Bo_d = 0$ and (b) $Bo_d = 10$. Other physical parameters are $Pe_s = 0.1$, $Ma_T = 0$, $\delta = 0.1$, $H = 10$, $e = 1$, $Ma_\Gamma = 0.1$, $Bo_s = 0$.

Now we look at the eastern hemispheres of the two droplets shown in figures 7(a) and (b). Keeping in mind that the primary cause of the lateral migration is the asymmetry in the surfactant concentration, we look into the extent of this asymmetry. A convenient way of representing this is by considering the quantity $(\Gamma_{max} - \Gamma_{min})$. A larger value of $(\Gamma_{max} - \Gamma_{min})$ would mean a greater amount of lateral migration of the droplet. On comparison of the extreme values of $\Gamma(\theta,\phi)$ we find that the value of $(\Gamma_{mas} - \Gamma_{min})$ is lower for the surfactant distribution depicted in figure 6(b). Therefore, an increase in the dilatational Boussinesq number results in the decrease in non-uniformity of surfactant distribution at the droplet interface. Therefore, the variations in the surface tension distribution would also be more uniform when the effects of interfacial viscosity are prominent which would lead to a decreased magnitude of the hydrodynamic forces at the droplet interface. Therefore, the lateral migration velocity of the droplet would also reduce because of the presence of interfacial viscosity.

5.3.2. *Non-isothermal Poiseuille flow: Temperature gradient applied in the flow direction*

Figure 8 depicts the variation in the cross-stream migration velocity of the droplet ($U_x$) with $\lambda$ for different values of the dilatational Boussinesq number ($Bo_d$). It can be observed that the magnitude of $U_x$ is much higher due to the presence of the temperature gradient along the direction of the flow. However, as the value of $Bo_d$ is increased, the magnitude of $U_x$ decreases. In order to explain the decrease in the magnitude of $U_x$, the distribution of surfactant concentration at the droplet interface is plotted for two different values of the dilatational Boussinesq number.

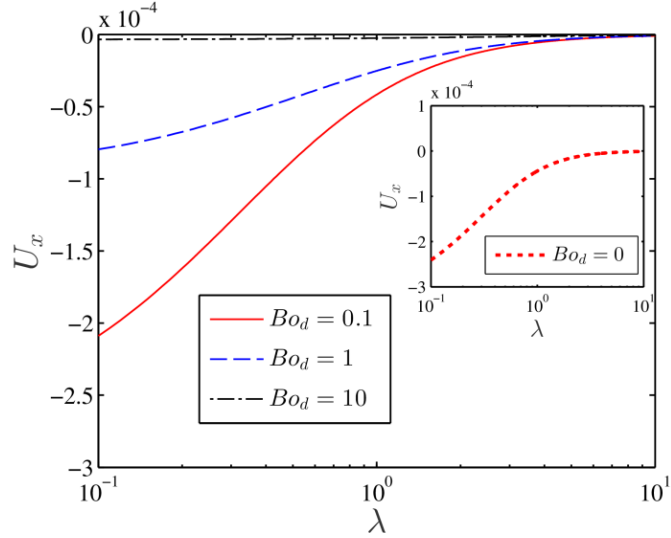

FIGURE 8. The variation of $U_x$ is plotted with $\lambda$ for various values of $Bo_d$ when $Ma_T = 0.1$, $e = 1$, $Pe_s = 0.1$, $Bo_s = 0$, $H = 10$, $Ma_\Gamma = 0.1$, $\delta = 0.1$ and $\zeta = 1$.

    In presence of temperature gradient along the direction of flow, the distribution of the surfactant concentration at droplet interface is shown in figure 9. There are several interesting features to note in the contour plots presented in figure 9. Firstly, as compared with the contour plots shown in figure 7, the lower hemispheres of the droplets in figure 9 are inactive as far as exhibiting non uniformities in surfactant concentration are concerned. This explains a greater magnitude of $U_x$ of the droplets when an external temperature field is applied in the direction of the flow. Now we compare figure 9(a) and 9(b). In order to provide physical justification to the decrease in magnitude of $U_x$, we again compare the values of $(\Gamma_{max} - \Gamma_{min})$ in both the cases. On doing so we conclude that the value of $(\Gamma_{max} - \Gamma_{min})$ is lower in case (b) and therefore the presence of interfacial viscosity serves to decrease the non-uniformity in the surfactant concentration at the interface which results in a lower magnitude of $U_x$.

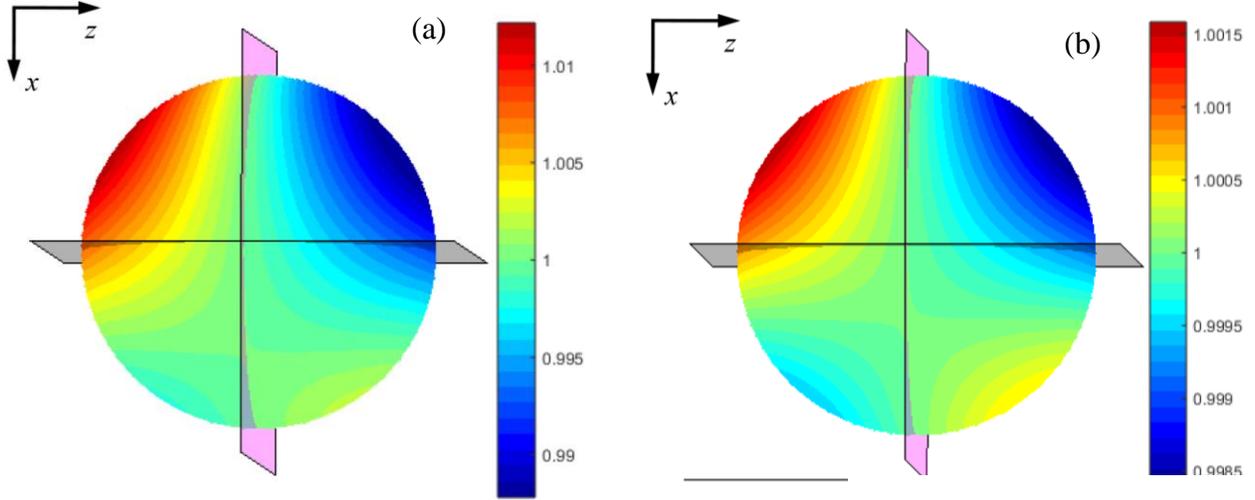

FIGURE 9. Surface plot of the surfactant concentration at the droplet interface for (a) $Bo_d = 0$ and (b) $Bo_d = 10$. Other physical parameters are $Pe_s = 0.1$, $Ma_T = 0.1$, $\zeta = 1$, $\delta = 0.1$, $H = 10$, $e = 1$, $Ma_\Gamma = 0.1$, $Bo_s = 0$.

Figure 10 shows the variation in $U_x$ with $\lambda$ for three different values of the shear Boussinesq number ($Bo_s$). We observe that as the value of $Bo_s$ is increased from 0.01 to 10, the magnitude of the lateral migration velocity decreases by a noticeable amount. However, as the magnitude of $Bo_s$ is increased further, we don't see any more significant changes in the magnitude of $U_x$.

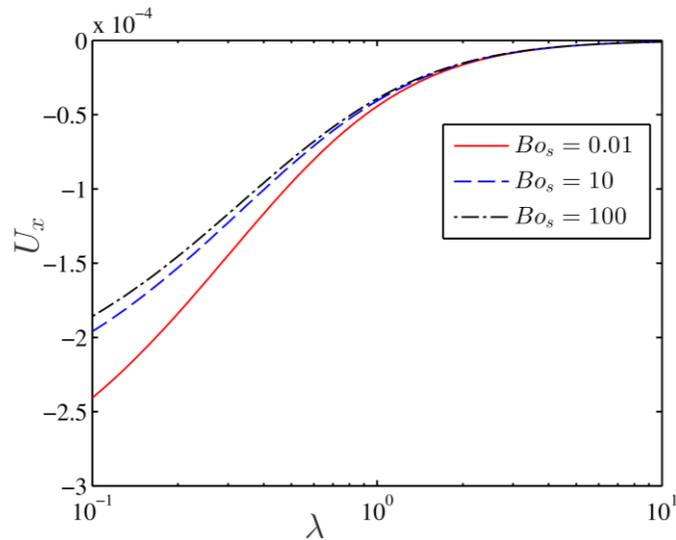

FIGURE 10. The variation of $U_x$ is plotted with the $\lambda$ for various values of $Bo_s$ when $Ma_T = 0.1$, $e = 1$, $Pe_s = 0.1$, $Bo_d = 0$, $H = 10$, $Ma_\Gamma = 0.1$, $\zeta = 1$ and $\delta = 0.1$.

5.3.3. *Non-isothermal Poiseuille flow: Temperature gradient opposite to the flow direction*

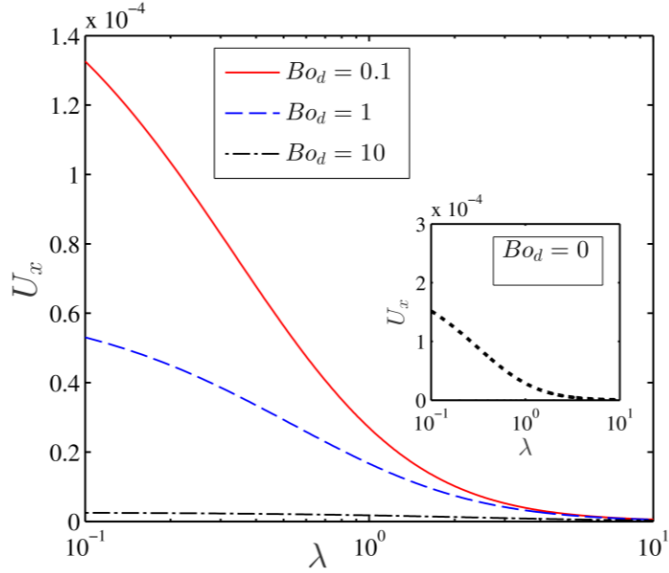

FIGURE 11. The variation of $U_x$ is plotted with $\lambda$ for various values of number $Bo_d$ when $Ma_T = 0.1$, $e = 1$, $Pe_s = 0.1$, $Bo_s = 0$, $H = 10$, $Ma_\Gamma = 0.1$, $\delta = 0.1$ and $\zeta = -1$.

Now we consider a case where the temperature gradient is applied opposite to the direction of the incipient flow. Towards this, we plot the variation of $U_x$ with $\lambda$ for different values of the dilatational Boussinesq number $Bo_d$. For the sake of comparison, at the inset of the figure, we have also inserted the plot of $U_x$ vs. $\lambda$ for $Bo_d = 0$. Firstly, it can be clearly noted that the direction of $U_x$ has reversed. Further, the magnitude of $U_x$ decreases as the value of $Bo_d$ is increased. In order to physically account for this phenomenon, we plot the distribution of surfactant concentration at the interface of the droplet. Figure 12 shows the contour plots for the surfactant concentration at the droplet interface. Here we see that the surfactant concentration appears to be reversed in comparison to figure 8 about the axial plane of the droplet. Due to the presence of a temperature gradient opposite to the direction of the imposed flow, the lower hemisphere of the droplet exhibits a greater non uniformity in the surfactant concentration than the upper hemisphere. Therefore, the droplet migrates in the opposite direction. Now on comparing the magnitudes of $(\Gamma_{max} - \Gamma_{min})$ between figures 12(a) and 12(b), we find that $(\Gamma_{max} - \Gamma_{min})$ is lower in figure 12(b). Therefore, the magnitude of the cross-stream migration velocity becomes lesser when the value of the dilatational Boussinesq number is increased.

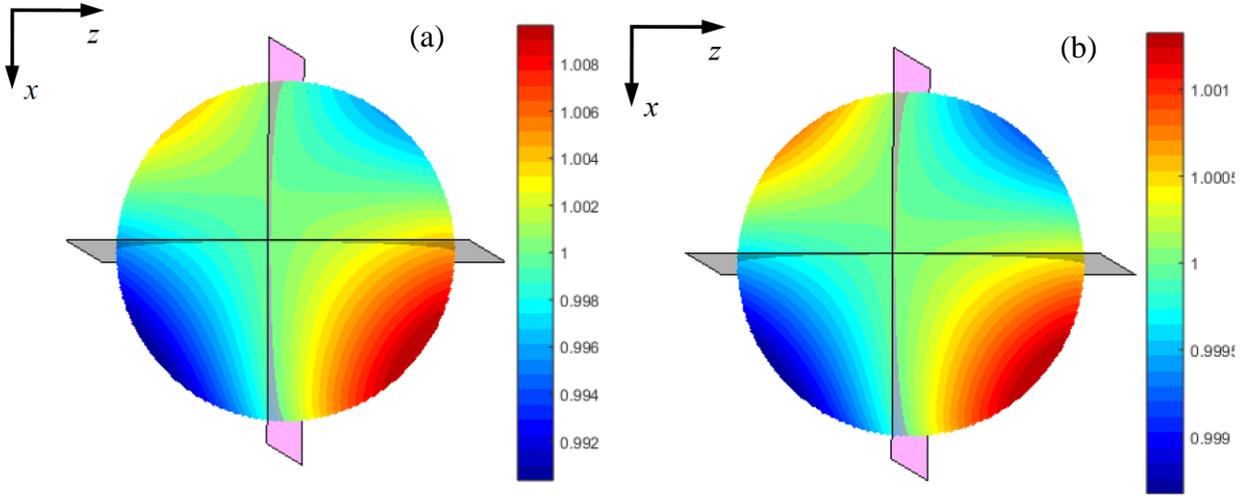

FIGURE 12. Surface plot of the surfactant concentration at the interface of the droplet for (a) $Bo_d = 0$ and (b) $Bo_d = 10$. Other physical parameters are $Pe_s = 0.1$, $Ma_T = 0.1$, $\zeta = -1$, $\delta = 0.1$, $H = 10$, $e = 1$, $Ma_\Gamma = 0.1$, $Bo_s = 0$.

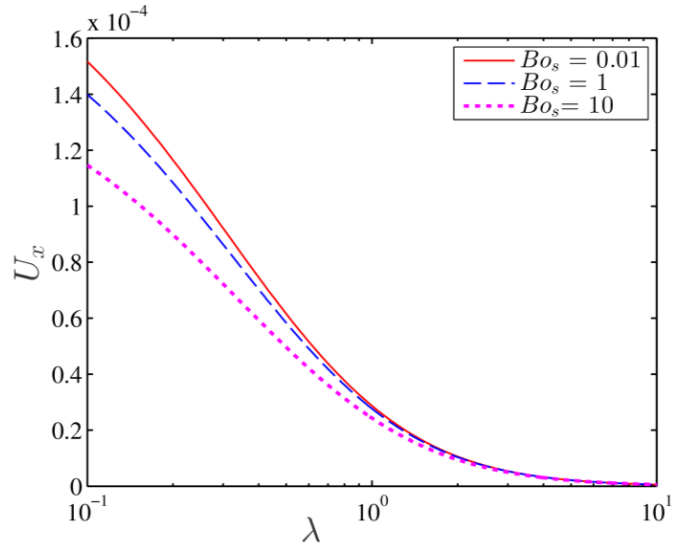

FIGURE 13. The lateral migration velocity of the droplet ($U_x$) is plotted with $\lambda$ for various values of $Bo_s$ when $Ma_T = 0.1$, $e = 1$, $Pe_s = 0.1$, $Bo_d = 0$, $H = 10$, $\delta = 0.1$ and $\zeta = -1$.

Figure 13 depicts the variation in the lateral migration velocity ($U_x$) with $\lambda$ for various values of the shear Boussinesq number ($Bo_s$). We note that on increasing the value of $Bo_s$, the magnitude of $U_x$ decreases. However, in this case the decrease is uniform in nature over the values of $Bo_s$ ranging from 0.01 to 10.

## 6. Conclusions

In the current analysis, a theoretical model is developed in order to investigate the influence of interfacial viscosity on the lateral migration of a surfactant-coated Newtonian droplet suspended in a non-isothermal pressure driven flow. Neglecting the effects of fluid inertia, shape deformation and thermal convection, we have obtained analytical expressions for the lateral migration velocity of the droplet for two limiting cases, namely: i) $Pe_s \ll 1$, which corresponds to the diffusion dominated transport of surfactants and ii) $Pe_s \gg 1$, which corresponds to the convection dominated surfactant transport. The first scenario is more relevant from an experimental perspective and therefore discussed in detail. In addition to that, we have also performed experiments in order to validate our theoretical findings and studied the effect of channel confinement on the lateral migration of the droplet. After analyzing all the important parameters involved in the physical problem, the following conclusions are made:

(i) The analytical solution considering the effect of interfacial viscosity shows a good agreement with the experimental result on lateral migration of a surfactant-coated droplet in a non-isothermal Poiseuille flow and provides a much greater approximation to the naturally occurring lateral migration of the droplet under same condition. Another important fact is that for identical values of the channel width and identical thermal condition, a larger droplet migrates faster towards the centerline of the flow as compared to a smaller one.

(ii) It has been found that the dilatational surface viscosity suppresses the lateral migration of the droplet to a large extent. The effect of dilatational interfacial viscosity has been investigated for three different situations namely i) isothermal flow, ii) non isothermal flow where the temperature gradient is along the direction of the flow and iii) non-isothermal flow having a uniform temperature gradient acting in a direction opposite to the incipient flow. In all the three cases, increase in the value of the dilatational Boussinesq number ($Bo_d$) results in decrease in the magnitude of the lateral migration velocity.

(iii) The shear surface viscosity arises also suppresses the lateral migration velocity of the droplet. Nevertheless, the decrease in the magnitude of lateral migration velocity due to shear surface viscosity is lesser as compared to the decrease occurring due to dilatational surface viscosity. Further, it has been noted that the nature of the decrease in the migration velocity due to shear viscosity is dependent on the thermal environment of the imposed flow. It has been noted that the decrease in the migration velocity due to shear surface viscosity is more gradual in presence of a uniform temperature gradient opposite to the direction of the flow as compared to an isothermal flow or a uniform temperature field along the flow direction.

**Supplementary material**

Supplementary material contains the details of constant coefficient of equation (18),(21), (26) and (33).